\documentclass[prb,preprint,superscriptaddress,pdf]{revtex4-1}
\usepackage{epsfig}
\usepackage{graphicx}
\usepackage{dcolumn}
\usepackage{bm}
\usepackage{tabularx}
\usepackage{hyperref}
\usepackage{xcolor}
\hypersetup{citecolor=black, filecolor= black, linkcolor= black, urlcolor= black}

\begin{document}

\preprint{version 0}

\title{Exploration of stable compounds, crystal structures, and superconductivity in the Be-H system}

\author{Shuyin Yu}
\affiliation{Science and Technology on Thermostructural Composite Materials Laboratory, School of Materials Science and Engineering, Northwestern Polytechnical University, Xi'an, Shaanxi 710072, PR China}
\author{Qingfeng Zeng}
\email{qfzeng@nwpu.edu.cn}
\affiliation{Science and Technology on Thermostructural Composite Materials Laboratory, School of Materials Science and Engineering, Northwestern Polytechnical University, Xi'an, Shaanxi 710072, PR China}
\author{Artem R. Oganov}
\affiliation{Department of Geosciences, Center for Materials by Design, and Institute for Advanced Computational Science, State University of New York, Stony Brook, NY 11794-2100, USA}
\affiliation{Moscow Institute of Physics and Technology, Dolgoprudny, Moscow Region 141700, Russia}
\affiliation{School of Materials Science and Engineering, Northwestern Polytechnical University, Xi'an, Shaanxi 710072, PR China}
\author{Chaohao Hu}
\affiliation{School of Materials Science and Engineering, Guilin University of Electronic Technology, Guilin 541004, People's Republic of China}
\author{Gilles Frapper}
\affiliation{IC2MP UMR 7285, Universit\'{e} de Poitiers - CNRS, Poitiers 86022, France}
\author{Litong Zhang}
\affiliation{Science and Technology on Thermostructural Composite Materials Laboratory, School of Materials Science and Engineering, Northwestern Polytechnical University, Xi'an, Shaanxi 710072, PR China}

\begin{abstract}
Using first-principles variable-composition evolutionary methodology, we explored the high-pressure structures of beryllium hydrides between 0 and 400 GPa. We found that BeH$_2$ remains the only stable compound in this pressure range. The pressure-induced transformations are predicted as $Ibam$ $\rightarrow $ $P\bar{3}m1$ $\rightarrow $ $R\bar{3}m$ $ \rightarrow $ $Cmcm$ $ \rightarrow $ $P4/nmm$, which occur at 24, 139, 204 and 349 GPa, respectively. $P\bar{3}m1$ and $R\bar{3}m$ structures are layered polytypes based on close packings of H atoms with Be atoms filling octahedral voids in alternating layers. $Cmcm$ and $P4/nmm$ contain two-dimensional triangular networks with each layer forming a kinked slab in the $ab$-plane. $P\bar{3}m1$ and $R\bar{3}m$ are semiconductors while $Cmcm$ and $P4/nmm$ are metallic. We have explored superconductivity of both metal phases, and found large electron-phonon coupling parameters of $ \lambda $=0.63 for $Cmcm$ with a $T_c$ of 32.1-44.1 K at 250 GPa and $ \lambda $=0.65 for $P4/nmm$ with a $T_c$ of 46.1-62.4 K at 400 GPa. The dependence of $T_c$ on pressure indicates that $T_c$ initially increases to a maximum of 45.1 K for $Cmcm$ at 275 GPa and 97.0 K for $P4/nmm$ at 365 GPa, and then decreases with increasing pressure for both phases.
\end{abstract}
\maketitle

\section{Introduction}

The search for new high-temperature superconductors has attracted great enthusiasm in both fundamental and applied research. Owing to its low mass and high electron density, \textquotedblleft metallic hydrogen$\textquotedblright$ has been predicted to possess a high superconducting transition temperature (\textit{T$_c$} $>$ 200 K)\cite{ashcroft1968metallic,barbee1989first,cudazzo2008ab}. However, hydrogen remains insulating at extremely high pressure ($>$ 320 GPa\cite{loubeyre2002optical}), which are too high for any applications. Another feasible method of obtaining the properties of metallic hydrogen is to form hydrogen-rich alloys with other elements\cite{ashcroft2004hydrogen}. Due to \textquotedblleft chemical precompression\textquotedblright, the pressure of metallization may be reduced significantly.

Inspired by the elusive state of matter, theoretical and experimental research has made considerable progress towards exploring superconductivity in hydrogen-rich compounds, e.g. for group IVa hydrides, calculations predicted that SiH$_4$\cite{feng2006structures,yao2007superconductivity,martinez2009novel}, GeH$_4$\cite{gao2008superconducting,li2007first}, SnH$_4$\cite{tse2007novel,gao2010high} and PbH$_4$\cite{zaleski2011high} may become superconductors at high (yet lower than pure H) pressure. The origin of high-pressure superconductivity can be derived from the particular \textquotedblleft H$_2$$\textquotedblright$ units, which are a feature common to hydrides of alkali metals \cite{zurek2009little}, alkaline earth metals \cite{hooper2013polyhydrides,hooper2014composition} and group IVa elements\cite{gao2008superconducting,gao2010high}. Experiments suggested metallization of SiH$_4$ at $\sim$60 GPa\cite{chen2008pressure} and its superconducting transition temperature (\textit{T$_c$}) is 17 K at 96 and 120 GPa\cite{eremets2008superconductivity}, though debates remain. In addition, the superconductivity of group IIIa hydrides (BH\cite{hu2013pressure}, AlH$_3$\cite{goncharenko2008pressure,islam2010alh} and GaH$_3$\cite{gao2011metallic}) and alkaline earth metal hydrides (CaH$_n$\cite{wang2012superconductive}, SrH$_n$\cite{hooper2014composition} and BaH$_n$\cite{hooper2013polyhydrides}) have also been extensively explored.

Beryllium hydrides can be an interesting subject of study, because low atomic mass of Be may lead to very high \textit{T$_c$} values. The only known beryllium hydride is BeH$_2$. The ground-state structure of BeH$_2$ is body-centered orthorhombic with $\textit{Ibam}$\cite{smith1988crystal} symmetry. At ambient conditions, BeH$_2$ is an insulator with a pronounced band gap of 5.5 eV\cite{wang2010mechanical}. Vajeeston et al.\cite{vajeeston2004structural} proposed that BeH$_2$ undergoes a series of phase transitions $\alpha$ $\rightarrow$ $\beta$ $\rightarrow$ $\gamma$ $\rightarrow$ $\delta$ $\rightarrow$ $\epsilon$ at pressures of 7, 51, 87 and 98 GPa, respectively, and reported that BeH$_2$ remains insulating up to 100 GPa. Zhang et al.\cite{zhang2010chemical} systematically investigated the pressure-induced metallization of alkaline earth hydrides, and found the metallization pressure of \textit{Pnma}-BeH$_2$ to be greater than 300 GPa. Wang et al.\cite{wang2014metallization} predicted that BeH$_2$ reaches a metallic state by a \textit{R$\bar{3}$m} $ \rightarrow $ \textit{Cmcm} phase transition, instead of a direct band gap closure in \textit{R$\bar{3}$m} phase.

\section{Computational methodology}

First-principles variable-composition evolutionary simulations were performed at 0, 50, 100, 150, 200, 250, 300 and 400 GPa using the USPEX code\cite{oganov2006crystal,oganov2011evolutionary,lyakhov2013new,oganov2010evolutionary}, which has the capability of discovering possible stoichiometries and the corresponding stable and metastable structures at given pressure-temperature conditions, and has successfully predicted a large number of stable structures\cite{zeng2013prediction,zhang2013unexpected,zhu2013novel}. The initial generation of structures and compositions was produced randomly with the use of space groups picked randomly from the total list of 230 groups. 50$ \% $ of the lowest-enthalpy structures were used as parents for the next generation. In addition, 20$\%$ of structures in each new generation were produced by lattice mutation, 15$\%$ by atomic transmutation and 15$\%$ were produced randomly. Each generation contained 60 structures and runs proceeded for up to 50 generations.

The underlying structure relaxations were carried out using the Vienna \textit{Ab-initio} Simulation Package (VASP) code\cite{kresse1996efficient}, in the framework of density functional theory (DFT)\cite{giannozzi1991ab,baroni1987green} within the Perdew Burke Ernzerhof generalized gradient approximation (PBE-GGA)\cite{perdew1996generalized}. The frozen all-electron projected augmented wave approach (PAW)\cite{blochl1994projector} was adopted to describe the core electrons and their effects on valence orbitals. A plane-wave kinetic energy cutoff of 600 eV and dense Monkhorst-Pack $\textit{k}  $-point grids\cite{monkhorst1976special} with a resolution higher than 2$\pi$$ \times $0.06 $ \AA{} $$^{-1}$ were used for all structures. The most stable structures were studied further at increased accuracy using a reciprocal-space grid better than 2$\pi$$ \times $0.03 $ \AA{} $$^{-1}$.

Phonon calculations were carried out using the supercell approach as implemented in the PHONOPY code\cite{togo2008first}. Electron-phonon coupling (EPC) calculations were explored using the pseudopotential plane-wave method within PBE-GGA, as implemented in the Quantum-Espresso package\cite{giannozzi2009quantum}. In these calculations, we used the kinetic energy cutoff of 60 Ry and Monkhorst-Pack $ \textit{k} $-point grids of 20$ \times $20$ \times $12 for the \textit{Cmcm} phase and 16$ \times $16$ \times $8 for the \textit{P4/nmm} phase with a Methfessel-Paxton\cite{methfessel1989high} smearing factor of 0.05 Ry. Additionally, \textit{q}-meshes of 5$ \times $5$ \times $3 for \textit{Cmcm} and 4$ \times $4$ \times $2 for \textit{P4/nmm} were used to calculate the electron-phonon coupling matrix elements, respectively. We used the Allen-Dynes-modified McMillan equation\cite{allen1975transition} to estimate \textit{T$ _{c} $}, as follows:
\begin{equation}
{T_c} = \frac{{{\omega _{\log }}}}
{{1.2}}\exp \left[ { - \frac{{1.04(1 + \lambda )}}
{{\lambda  - {\mu ^*}(1 + 0.62\lambda )}}} \right]
\end{equation}
where $ \omega $$ _{log} $ is the logarithmic average frequency, $ \lambda $ is the electron-phonon coupling constant and $ \mu $$ ^{\ast} $ is the Coulomb pseudopotential, which is assumed to be between 0.10-0.13\cite{ashcroft2004hydrogen}.

\section{Results and discussions}

\begin{figure}[htbp]
\includegraphics[width=0.5\linewidth]{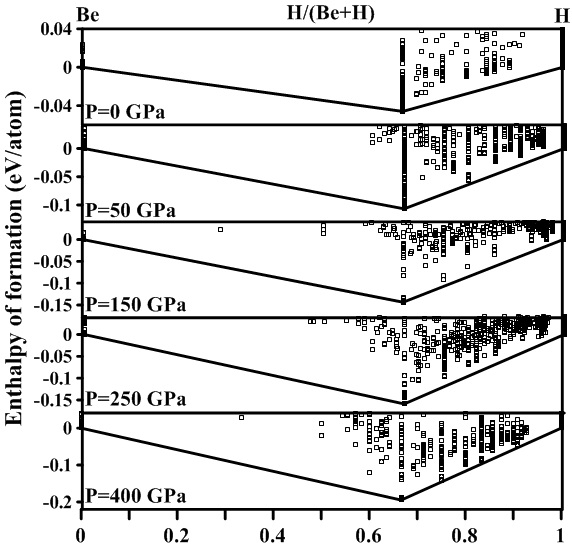}
\caption{\label{Fig1} (Color online). Convex hull phase diagrams for the Be-H system at 0, 50, 150, 250 and 400 GPa.}
\end{figure}
Fig. \ref{Fig1} shows the convex hull phase diagrams for the Be-H system at selected pressures. The ground-state enthalpy of formation $\Delta$H$_{f}$ is defined as $\Delta$H$_{f}$(Be$ _{x} $H$_{y}$)=$\Delta$H(Be$ _{x} $H$_{y}$) - x$\Delta$H(Be) - y$\Delta$H(H). A compound is thermodynamically stable if it has lower enthalpy than any isochemical mixture of the elements or other compounds. Such stable compounds form the convex hull. Based on our evolutionary searches, elemental Be adopts the $ \textit{P6$ _{3} $/mmc} $ structure below 390 GPa and $ \textit{bcc Im$\bar{3}$m}$ structure above 390 GPa. Our findings are in good agreement with previous calculations\cite{guo2014phase}. Hydrogen undergoes a series of phase transitions: $ \textit{P6$ _{3} $/m} $ (P $<$ 105 GPa), $ \textit{C2/c} $ (105 $<$ P $<$ 207 GPa), $ \textit{Cmca-12} $ (270 $<$ P $<$ 385 GPa), \textit{Cmca} (P $> $ 385 GPa)\cite{pickard2007structure}, in addition to the experimentally known \textit{Ibam}\cite{smith1988crystal} structure, we found a series of pressure-induced structural transformations (\textit{Ibam} $\rightarrow $ \textit{P$\bar{3}$m1} $\rightarrow $ \textit{R$\bar{3}$m} $ \rightarrow $ \textit{Cmcm} $ \rightarrow $ \textit{P4/nmm}) with increasing pressure. Notably, we did not find other stable compounds besides BeH$_{2} $ over the entire pressure range 0 - 400 GPa. The detailed structural parameters of these predicted phases are summarized in Tab. \ref{Tab1}.

\begin{figure}[htbp]
\includegraphics[width=0.6\linewidth]{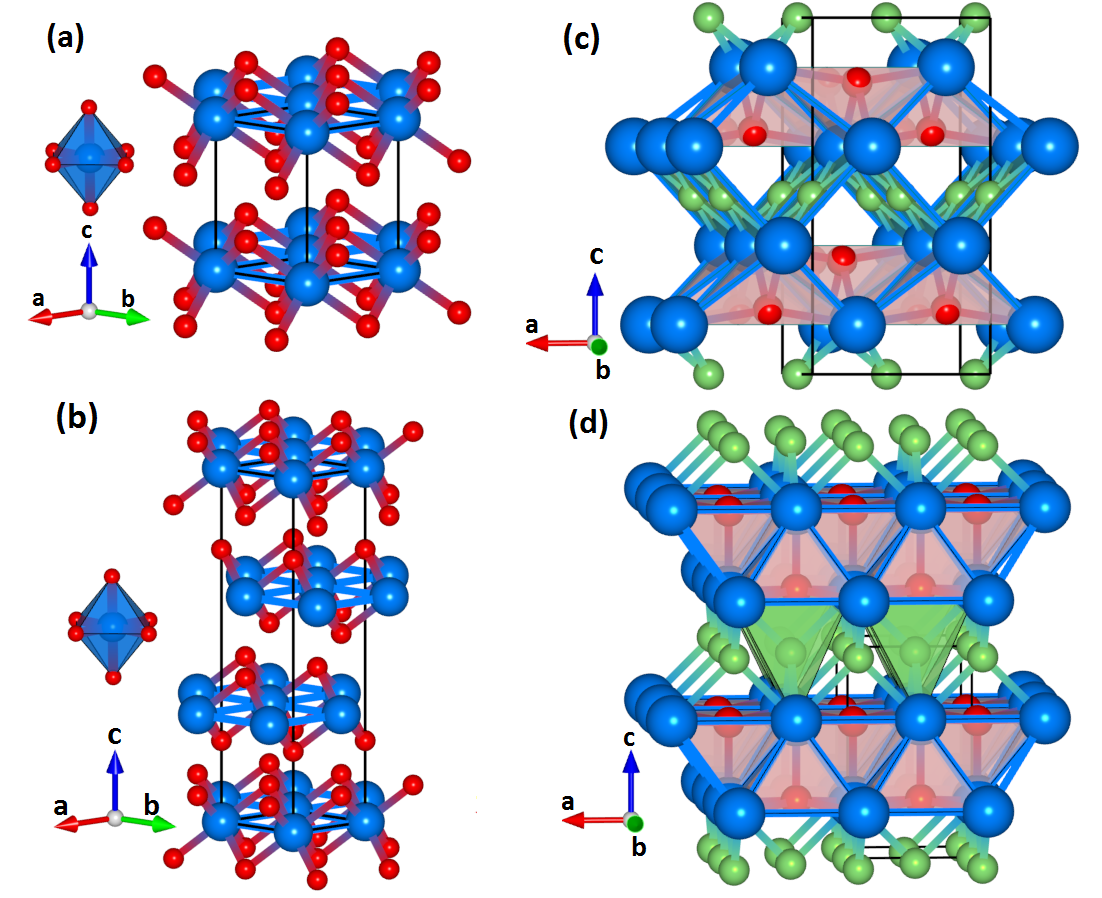}
\caption{\label{Fig2} (Color online). Extended crystal structures of solid BeH$_{2} $ for (a) the $\textit{P$\bar{3}$m1}$ structure at 50 GPa; (b) the $\textit{R$\bar{3}$m}$ structure at 150 GPa; (c) the \textit{Cmcm} structure at 250 GPa; and (d) the \textit{P4/nmm} structure at 400 GPa. The large blue spheres represent Be atoms, while the small red and green spheres indicate two  symmetrically inequivalent H atoms.}
\end{figure}

\begin{figure}[htbp]
\includegraphics[width=1\linewidth]{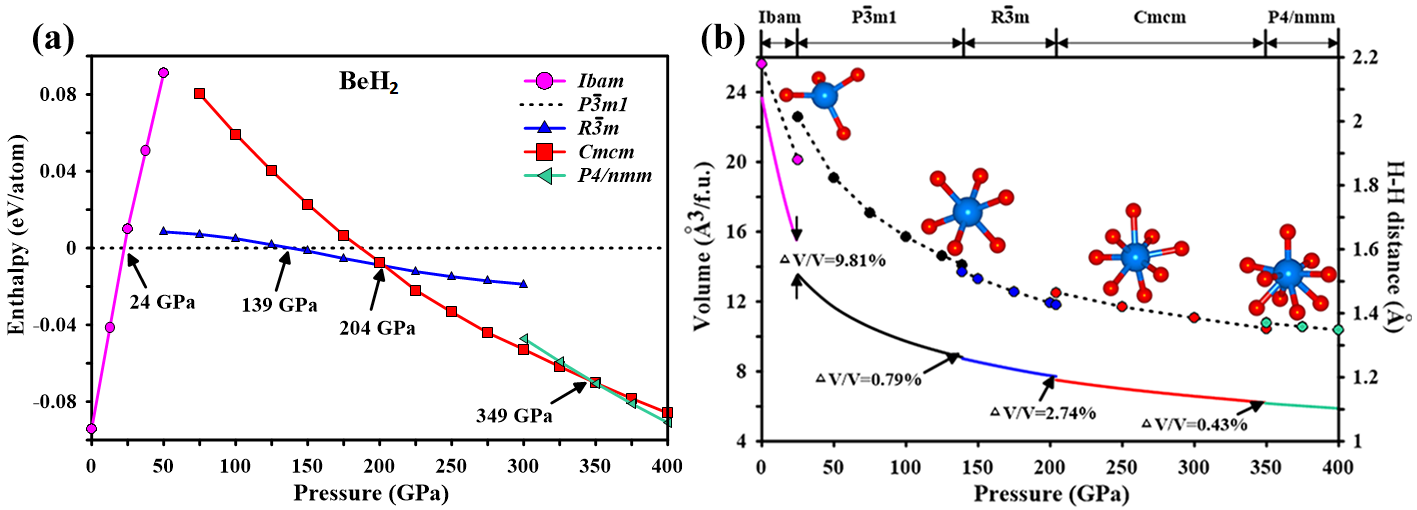}
\caption{\label{Fig3} (Color online). (a) Enthalpy per atom for various BeH$_{2} $ structures as a function of pressure with the $\textit{P$\bar{3}$m1}$ structure taken as the reference; (b) Computed equations of state of BeH$_2$ (solid lines) and shortest H-H distance (dotted lines).}
\end{figure}

At ambient conditions, BeH$_{2} $ crystallizes in the orthorhombic \textit{Ibam} structure. This structure consists of a three-dimensional network of distorted tetrahedra with Be atoms sitting at the center of the tetrahedra and H atoms at the corner in a bridged position between two Be atoms. The orthorhombic phase transforms to a CdI$_{2} $-type structure ($\textit{P$\bar{3}$m1}$; Fig. \ref{Fig2}a) at 24 GPa, then it transforms to a related CdCl$_{2} $-type structure ($\textit{R$\bar{3}$m}$; Fig. \ref{Fig2}b) at 139 GPa (Fig. \ref{Fig3}a). Both structures are made of layers of edge-sharing BeH$_{6}$-octahedra, but stacking sequences of these layers are different. The shortest interlayer H-H distances decrease from 1.83 \AA{} at 50 GPa to 1.51 \AA{} at 150 GPa (Fig. \ref{Fig3}b).

\begin{table*}[htbp]
\centering
\caption{\label{Tab1} Optimized structural parameters for the predicted BeH$ _{2} $ structures at selected pressures.}
\begin{tabular}{cccclllll}
\hline
\hline
Pressure & Space group& No.& Lattice parameters & Atom& \multicolumn{4}{c}{Wyckoff positions}\\
\cline{6-9}
   (GPa) &       &                     &    (\AA{}, deg)    &                      &  Sites  & x & y & z \\
\hline
50 & $\textit{P$\bar{3}$m1}$& 164& \textit{a} = 2.085  & Be& \textit{1a} & 0& 0& 0\\
   &                        &    & \textit{c} = 3.104  &  H& \textit{2d} & 0.667& 0.333& 0.721\\
150& $\textit{R$\bar{3}$m}$ & 166& \textit{a} = 1.886  & Be& \textit{1b} & 0& 0& 0.5\\
   &                        &    & \textit{c} = 8.316  &  H& \textit{2c} & 0& 0& 0.732\\
250& \textit{Cmcm}          & 63 & \textit{a} = 1.796  & Be& \textit{4c} & 0& 0.139& 0.75\\
   &                        &    & \textit{b} = 5.503  & H1& \textit{4b} & 0& 0.5& 1\\
   &                        &    & \textit{c} = 2.840  & H2& \textit{4c} & 0& 0.819& 0.75\\
400& \textit{P4/nmm}        & 129& \textit{a} = 1.906  & Be& \textit{2c} & 0.5& 0& 0.707\\
   &                        &    & \textit{c} = 3.240  & H1& \textit{2a} & 0& 0& 0\\
   &                        &    &                     & H2& \textit{2c} & 0.5& 0& 0.327\\
\hline
\hline
\end{tabular}
\end{table*}

We found a similar high-pressure structure in the B-H system\cite{hu2013pressure}: at P $>$ 168 GPa, the \textit{P6/mmm}-BH structure is the most stable phase, and may be described as stacking BH-layers with planar closely-packed arrays of boron atoms. On-top H atoms locate symmetrically between the boron layers. Note that if one assigns a formal charge of -1 to H (hydride-like), both the boron atoms in BH and beryllium atoms in BeH$_{2} $ have a formal ns$^2$ valence electron configuration. Notably, the structural transformation from \textit{Ibam} to $\textit{P$\bar{3}$m1}$ is accompanied by a large density jump of 9.81$\%$ while only 0.79$\%$ increase occurs at the transition from $\textit{P$\bar{3}$m1}$ to $\textit{R$\bar{3}$m}$ (Fig. \ref{Fig3}b). The structures in ref. \cite{smith1988crystal} are metastable with respect to the $\textit{P$\bar{3}$m1}$ structure.

At 204 GPa, the layered $\textit{R$\bar{3}$m}$ structure transforms into an orthorhombic \textit{Cmcm} structure (Fig. \ref{Fig2}c). In this structure, Be atoms are coordinated by eight hydrogens, whereas hydrogens are in the fourfold coordination (H1 atoms - in planar square coordination, H2 atoms - in tetrahedral coordination). Note that H1 atoms form flat pure-hydrogen rectangular layers with shortest H-H distance of 1.42 \AA{} at 250 GPa. The tetragonal \textit{P4/nmm} structure (PbClF-type) becomes stable at 349 GPa. In this structure, Be atoms are coordinated by nine hydrogen atoms (forming a capped tetragonal antiprism); H1 atoms are in a fourfold (planar square) coordination and H2 atoms in a fivefold (square pyramid) coordination. The structure can be viewed as layered, with double layers formed by Be and H2 atoms, alternating with square layers formed by H1 atoms. The Be-Be distances at 400 GPa are 1.901 and 1.906 \AA{} within the double layers, and 2.328 \AA{} between these layers, reinforcing cohesion of the highly delocalized covalent three-dimensional BeH$_2$ structure. The shortest H-H distance is between H1 atoms, i.e. in the pure-hydrogen square layer - at 400 GPa this distance is 1.345 \AA{}. At such distances overlap of atomic orbitals is strong enough to make the material metallic. This is very similar to the \textit{Pbcn}-SiH$_{4} $\cite{martinez2009novel} with the closest H-H distance of 1.35 \AA{}.

\begin{figure}[htbp]
\includegraphics[width=0.6\linewidth]{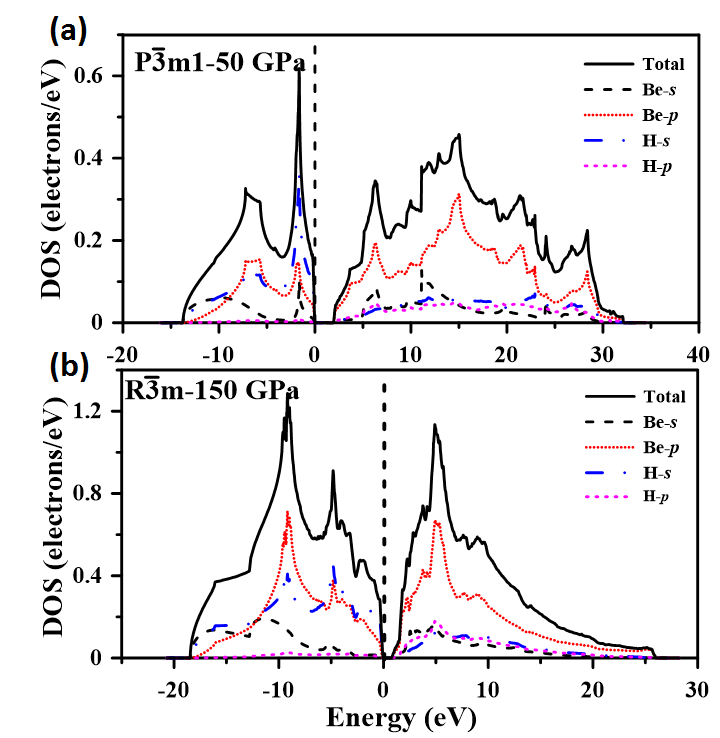}
\caption{\label{Fig4} (Color online). Total and Partial density of states (DOS) for the (a) $\textit{P$\bar{3}$m1}$ phase at 50 GPa and (b) \textit{R$\bar{3}$m} phase at 150 GPa.}

\end{figure}

Fig. \ref{Fig4} shows electronic densities of states (DOS) for the $\textit{P$\bar{3}$m1}$ and \textit{R$\bar{3}$m} structures, from which it can be clearly seen that both structures are semiconductors with band gaps of 1.95 eV and 0.74 eV, respectively. It is clear that there is substantial hybridization of the Be-\textit{p} states and H-\textit{s} states, suggesting large degree of covalency. The covalent bonds are mainly from the intralayer BeH$_{6} $ octahedra while the interlayer interactions are mainly van der Waals forces. Even at very high pressures these layered structures remain insulating (e.g., the \textit{R$\bar{3}$m} structure has band gap 0.18 eV at 200 GPa).

\begin{figure}[htbp]
\includegraphics[width=0.6\linewidth]{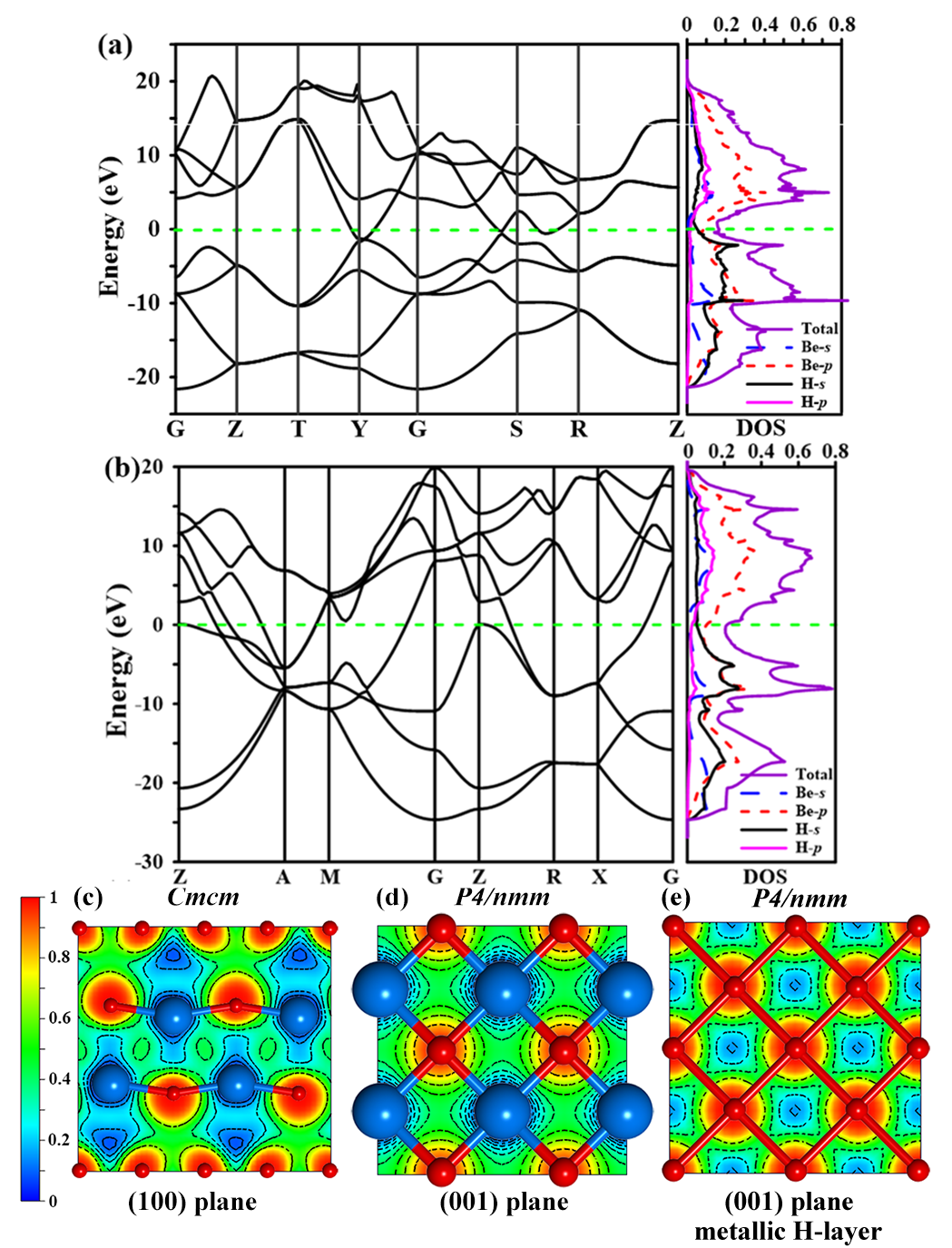}
\caption{\label{Fig5} (Color online). (a) and (b) band structures and partial DOS for \textit{Cmcm} phase at 250 GPa and \textit{P4/nmm} phase at 400 GPa, respectively; (c), (d) and (e) electron localization function (ELF) through specific surfaces.}
\end{figure}

Fig. \ref{Fig5} shows the band structures, partial DOS and electron localization function (ELF) for the \textit{Cmcm} phase at 250 GPa and \textit{P4/nmm} phase at 400 GPa, respectively. The band structures reveal that both structures are metallic with several bands crossing the Fermi level, and a pseudogap. The dispersed valence and conduction bands near the Fermi level signify a relatively large DOS at the Fermi level (0.098 and 0.107 electrons/eV/f.u., respectively), which may favor superconducting behavior. The valence band widths are greater than in the low-pressure phases, which indicates enhanced electron delocalization, thus, more electrons participate in bonding interactions, which promotes structural stability.

\begin{figure}[htbp]
\includegraphics[width=0.8\linewidth]{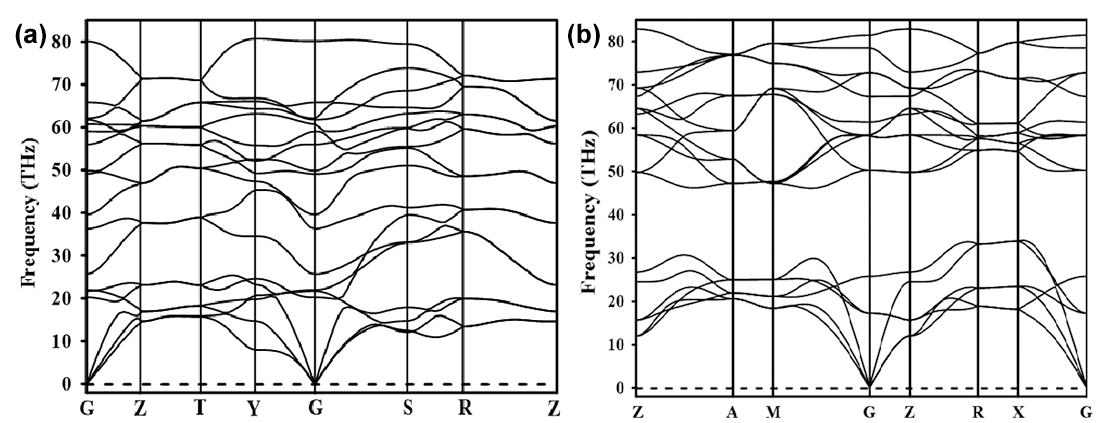}
\caption{\label{Fig6} (Color online). Calculated phonon spectra along selected high symmetry points for (a) the \textit{Cmcm} phase at 250 GPa and (b) \textit{P4/nmm} phase at 400 GPa.}
\end{figure}

Distributions of the electron localization function (ELF) reveal electron accumulation on the H atoms. For the \textit{P4/nmm} phase, the ELF between H atoms in the square H-layer is close to 0.5 (Fig. \ref{Fig5}e), equal to the value for the electron gas. The DOS and ELF analysis are also in agreement with Bader\cite{bader1990atoms} analysis. In the \textit{Cmcm} phase, Be atoms have charge +1.58, while the charges of H1 and H2 atoms are -0.82 and -0.76, respectively. In the \textit{P4/nmm} phase, the charges of Be, H1 and H2 atoms are +1.57, -0.94 and -0.63, respectively.

\begin{figure}[htbp]
\includegraphics[width=0.5\linewidth]{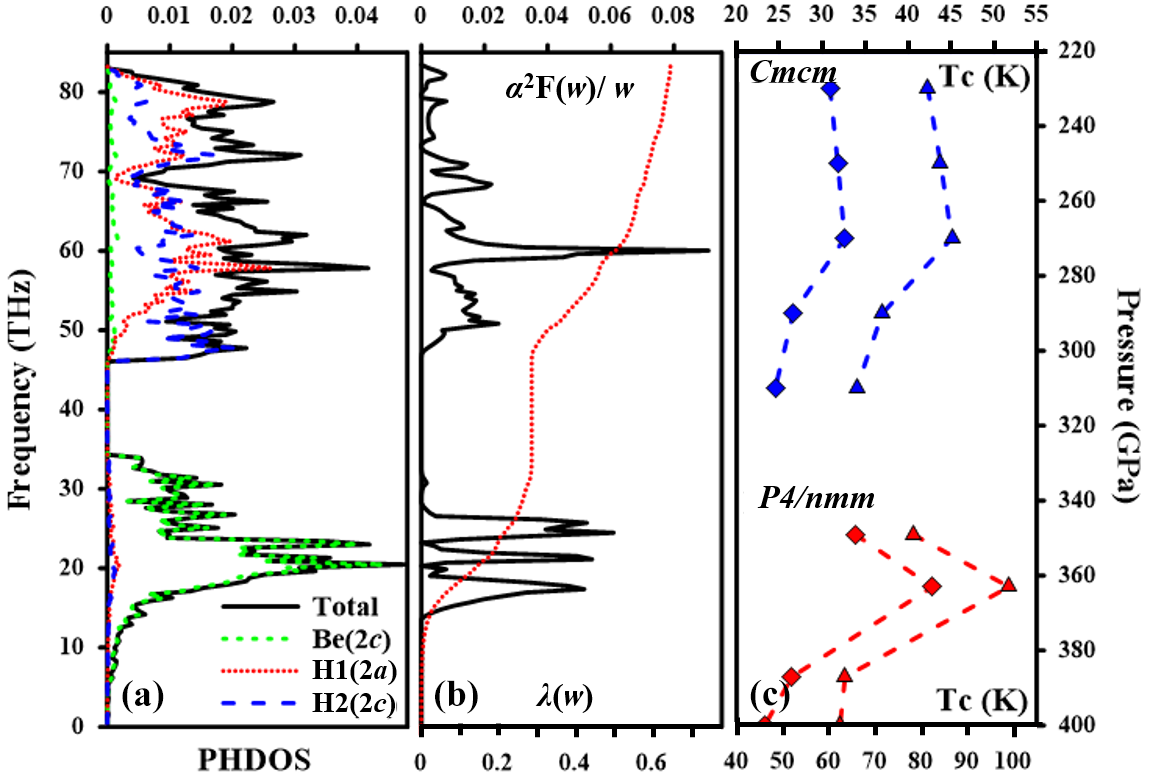}
\caption{\label{Fig7} (Color online). (a) Total and partial phonon density of states (PDOS) for \textit{P4/nmm} phase; (b) Eliashberg phonon spectral function $ \alpha $$ ^{2} $F($ \omega $) and electron-phonon integral $ \lambda $($ \omega $) as a function of frequency at 400 GPa; and (c) calculated superconducting transition temperature (\textit{T$ _{c} $}) \textit{vs} pressure for \textit{Cmcm} and \textit{P4/nmm} phases, triangles and squares represent $ \mu $$ ^{*} $=0.1 and 0.13, respectively.}
\end{figure}

The calculated phonon spectra for the \textit{Cmcm} and \textit{P4/nmm} phases establish dynamical stability, as there are no imaginary phonon frequencies anywhere in the Brillouin zone (Fig. \ref{Fig6}). We further explored the superconductivity for both structures by performing electron-phonon coupling (EPC) calculations. For the \textit{Cmcm} structure at 250 GPa, the electron-phonon coupling parameter $ \lambda $ is 0.63, indicative of quite strong EPC. Using the calculated $ \omega $$ _{log} $ of 1670.7 K and the commonly accepted values of the Coulomb pseudopotential  $ \mu $$ ^{*} $ (0.1-0.13)\cite{ashcroft2004hydrogen}, we obtained \textit{T$ _{c} $} in the range of 32.1-44.1 K using the modified Allen-Dynes-modified McMillan equation\cite{allen1975transition}. Our results are similar to those of Wang et al.\cite{wang2014metallization}, the differences being due to different pseudopotentials. From Fig. \ref{Fig7}c, it is clear that \textit{T$ _{c} $} of the \textit{Cmcm} structure first increases and then decreases with increasing pressure, and reaches a maximum (45.1 K) at $\sim$275 GPa.

Fig. \ref{Fig7} shows the total and partial phonon density of states together with the Eliashberg phonon spectral function $ \alpha $$ ^{2} $F($ \omega $) and electron-phonon integral $ \lambda $($ \omega $) as a function of frequency for the \textit{P4/nmm} structure at 400 GPa. Low-frequency ($ <$ 35 THz) vibrations are mostly related to Be atoms, while higher-frequency ($>$ 45 THz) modes mainly come from the vibration of H1 and H2 atoms. At 400 GPa, The calculated EPC parameter $ \lambda $ is 0.65, indicating  rather strong EPC in the \textit{P4/nmm} structure. Using the calculated $ \omega $$ _{log} $ of 2170 K and $ \mu $$ ^{*} $ (0.1-0.13), we obtained \textit{T$ _{c} $} in the range of 46.1-62.4 K. The vibrations of Be below 35 THz contribute about 44.3$\%$ of total $ \lambda $, while the vibrations of H1 and H2 above 45 THz contribute about 55.7$\%$ with no obvious difference between H1 and H2 vibrations to $ \lambda $. In addition, the pressure dependence of \textit{T$ _{c} $} displays the same trend as observed in the \textit{Cmcm} phase and reaches a maximum of 97.0 K at 365 GPa. This is one of the highest \textit{T$ _{c} $} values predicted in literature. Note that the Allen-Dynes formula is expected to be reliable when ¦Ë is less than 1-1.5\cite{allen1983theory}, which is the case here.

\section{Conclusions}
In summary, using variable-composition evolutionary simulations for crystal structure prediction, we investigated the high-pressure phases of solid beryllium hydrides in the pressure range of 0-400 GPa. BeH$_2$ is found to be the only stable beryllium hydride. The pressure-induced transformations are predicted to be \textit{Ibam} $\rightarrow $ \textit{P$\bar{3}$m1} $\rightarrow $ \textit{R$\bar{3}$m} $ \rightarrow $ \textit{Cmcm} $ \rightarrow $ \textit{P4/nmm}, which occur at 24, 139, 204 and 349 GPa, respectively. The layered \textit{P$\bar{3}$m1} and \textit{R$\bar{3}$m} structures belong to the well-known CdI$_2$ and CdCl$_2$ types, respectively. The \textit{Cmcm} and \textit{P4/nmm} phases contain 8- and 9-coordinate Be atoms, respectively, and layers of H atoms with short H-H distances, responsible for metallic conductivity. The entire phase transformations are first-order with volume shrinkage values of 9.81$\%$, 0.79$\%$, 2.71$\%$ and 0.43$\%$, respectively. The \textit{P$\bar{3}$m1} and \textit{R$\bar{3}$m} structures are semiconductors while the \textit{Cmcm} and \textit{P4/nmm} phases are metallic. Electron-phonon coupling calculations show that the \textit{Cmcm} and \textit{P4/nmm} structures are phonon-mediated superconductors, with large electron-phonon coupling parameters of 0.63 for the \textit{Cmcm} phase with a \textit{T$ _{c} $} of 32.1-44.1 K at 250 GPa and 0.65 for the \textit{P4/nmm} phase with a \textit{T$ _{c} $} of 46.1-62.4 K at 400 GPa. Dependence of \textit{T$ _{c} $} on pressure indicates that \textit{T$ _{c} $} will increase initially to a maximum value of 45.1 K for the \textit{Cmcm} phase at 275 GPa and 97.0 K for the \textit{P4/nmm} phase at 365 GPa, respectively, and then decrease with increasing pressure for both structures.

\begin{acknowledgments}
We thank the Natural Science Foundation of China (Grants No. 51372203, No. 11164005 and No. 51332004), the National Basic Research Program of China (973 Program, Grant No. 2014CB643703), the Basic Research Foundation of NWPU (Grant No. JCY20130114), the Foreign Talents Introduction and Academic Exchange Program (Grant No. B08040), the National Science Foundation (Grants No. EAR-1114313 and No. DMR-1231586), DARPA (Grants No. W31P4Q1310005 and No. W31P4Q1210008), and the Government of the Russian Federation (Grant No. 14.A12.31.0003) for financial support. The authors also acknowledge the High Performance Computing Center of NWPU, Shanghai Supercomputer Centre, and the National Supercomputing Center in Shenzhen and the GENCI-CINES (France) for the allocation of computing time on their machines.
\end{acknowledgments}

\bibliographystyle{apsrev4-1}
\bibliography{Reference}

\end{document}